\documentstyle[12pt]{article} 
 
\begin{document} 
\title{Meissner effect for color superconducting\\ quark matter} 
\author{D.M. Sedrakian$^1$, D. Blaschke$^2$,  K.M. Shahabasyan$^1$\\ 
D.N. Voskresensky$^3$\\ [5mm] 
$^1$Physics Department, Yerevan State University\\ 
Alex Manoogian str. 1, 375025 Yerevan, Armenia\\ 
$^2$Fachbereich Physik, Universit\"at Rostock\\ 
Universit\"atsplatz 1, D-18051 Rostock, Germany\\ 
$^3$Moscow Institute for Physics and Engineering\\ 
Kashirskoe shosse 31, 115409 Moscow, Russia} 
\date{ } 
\maketitle 
 
\begin{abstract} 
The behaviour of the magnetic field inside the superconducting quark matter 
core of a neutron star is investigated in the framework of the Ginzburg-Landau 
theory. 
We take into account the simultaneous coupling of the diquark condensate field 
to the usual magnetic and to the gluomagnetic gauge fields. 
We solve the problem for three different physical situations: a semi-infinite 
region with a planar boundary, a spherical region, and a cylindrical region. 
We show that Meissner currents near the quark core boundary effectively screen 
the external static magnetic field. 
\end{abstract} 
\vspace{3cm}

\noindent
MPG-VT-UR 213/00\\
December 2000 
\newpage
 
\section{ Introduction} 
 
Recently, possible formation of diquark condensates in QCD at finite density 
has been re-investigated in  series of papers following Refs.\cite{1},\cite{2}.
It has been shown that in chiral quark models with non-perturbative 4-point 
interaction motivated from instantons \cite{3} or non-perturbative gluon 
propagators \cite{4}, \cite{5} the anomalous quark pair amplitudes in the 
color antitriplet channel can be very large: of the order of 100 MeV. 
Therefore, one expects the diquark condensate to dominate the physics at 
densities beyond the deconfinement/chiral restoration transition density and 
below the critical temperature (of the order of 50 MeV). 
Various phases are possible. 
The so called two-flavor (2SC) and three-flavor (3SC) phases allow for 
unpaired quarks of one color. 
It has been also found \cite{6}, \cite{7} that there can exist a color-flavor 
locked (CFL) phase for not too large strange quark masses \cite{8}, where 
color superconductivity is complete in the sense that diquark 
condensation results in a pairing gap for the quarks of all three flavors and 
colors. 

The high-density phases of QCD at low temperatures are relevant for the 
explanation of phenomena in rotating massive compact stars which might 
manifest themselves as pulsars. 
Physical properties of these objects (once being measured) could constrain our 
hypotheses about the state of matter at the extremes of densities. 
In contrast to the situation for the cooling behaviour of compact stars 
\cite{9}, \cite{9a}, where the CFL phase is dramatically different from the 
2SC and 3SC phases, we don't expect qualitative changes of the magnetic field 
structure for these  phases. 
Therefore, below we shall restrict ourselves to the discussion of the simpler 
two-flavor theory first. 
Bailin and Love \cite{10} used a perturbative gluon propagator which yielded a 
very small pairing gap and they concluded that quark matter is a type I 
superconductor, which expells the magnetic field of a neutron star within 
time-scales of 10$^4$ years. 
If their arguments would hold in general, the observation of life-times for 
magnetic fields as large as 10$^7$years \cite{11}, \cite{12} would exclude the 
occurence of an extended superconducting quark matter core in pulsars. 
These estimates are not valid for the case of diquark condensates characterized
by large quark gaps. 
Besides, in Ref. \cite{13} the authors found that within recent 
non-perturbative approaches for the effective quark interaction that allow 
for large pairing gaps the quark condensate forms a type II superconductor. 
Consequently for the magnetic field $H<H_{c1}$ there exists a Meissner effect 
and for $H_{c2}>H>H_{c1}$ the magnetic field can penetrate into the quark core 
in quantized flux tubes. 
However, they have not considered in that paper the simultaneous coupling of 
the quark fields to the magnetic and gluomagnetic gauge fields. 
 
Though color and ordinary electromagnetism are broken in a color 
superconductor, there is a linear combination of the photon and the gluon that 
remains massless. 
The authors of Ref. \cite{14} have considered the problem of the presence of 
magnetic fields inside color superconducting quark matter taking into 
account the possibility of the so called ''rotated electromagnetism''. 
They came to the conclusion that there is no Meissner effect and  the external 
static homogeneous magnetic field can penetrate into superconducting quark 
matter because in their case it obeys the sourceless Maxwell equations. 
To our opinion this result is obtained when one does not pose correct boundary 
conditions for the fields. 
Obviously it is energetically favorable to expell the magnetic field rather 
than to allow its penetration inside the superconducting matter. 
Using for the description of the diquark condensate interacting with two 
gauge fields the same model as in Refs. \cite{9}, \cite{8} and \cite{15}, 
we will show that the presence of the massless excitation in the spectrum does 
not prevent the Meissner currents to effectively screen  the static external 
magnetic field. 
 
In Ref. \cite{15} two of us have derived the Ginzburg-Landau equations 
of motion for the diquark condensate 
placed in static magnetic and gluomagnetic fields, 
\begin{equation} 
ad_p+\beta (d_pd_p^{*})d_p+\gamma (i\bigtriangledown -\frac e3
\stackrel{\rightarrow }{A}+\frac g{\sqrt{3}}\stackrel{\rightarrow }{G}_8)^2d_p
=0, 
\label{ku} 
\end{equation} 
where $d_p$ is the order parameter, 
$a=t$ $dn/dE,\beta =(dn/dE)$ $7$ $\zeta (3)(\pi T_c)^{-2}/8,$ 
$\gamma =p_F^2$ $\beta /(6\mu ^2),$ $dn/dE=p_F$ $\mu /\pi ^2$, 
$t=(T-T_c)/T_c$, $T_c$ being the critical temperature, $p_F$-the quark 
Fermi momentum, and for the gauge fields 
\begin{equation} 
\lambda _q^2{\rm rot}~{\rm rot}\stackrel{\rightarrow }{A}+\sin 
^2\alpha \stackrel{\rightarrow }{A}=i\frac{\sin \alpha (d_p\nabla 
d_p^{*}-d_p^{*}\nabla d_p)}{2qd^2}+\sin \alpha \cos \alpha 
\stackrel{\rightarrow }{G}_{8,}  \label{ka} 
\end{equation} 
\begin{equation} 
\lambda _q^2{\rm rot}~{\rm rot}\stackrel{\rightarrow }{G}_8+\cos 
^2\alpha \stackrel{\rightarrow }{G}_8=-i\frac{\cos \alpha (d_p\nabla 
d_p^{*}-d_p^{*}\nabla d_p)}{2qd^2}+\sin \alpha \cos \alpha 
\stackrel{\rightarrow }{A}.  \label{kq} 
\end{equation} 
These equations introduce a ''new'' charge of the diquark pair 
$q=\sqrt{\eta^2e^2+g^2}P_8$, $P_8=1/\sqrt{3}$, 
and for the diquark condensate with paired blue-green and 
green-blue $ud$ quarks  one has $\eta =1/\sqrt{3}$. 
The penetration depth of the magnetic and gluomagnetic fields 
$\lambda _q$ and the mixing angle $\alpha $ are given by 
\begin{equation} 
\lambda _q^{-1}=qd \sqrt{2\gamma },\;\,\cos \alpha =\frac g{\sqrt{\eta 
^2e^2+g^2}}~.  \label{kb} 
\end{equation} 
At neutron star densities gluons are strongly coupled ($g^2/4\pi \sim 1)$ 
whereas photons are weakly coupled $(e^2/4\pi =1/137)$, 
so that $\alpha 
\simeq \eta e/g$ is small. 
For  $g^2/4\pi \simeq 1$ we get $\alpha \simeq 1/20$. 
The new charge $q$ is by an order of magnitude  larger than $e/\sqrt{3}$. 
 
Please notice also that since red quarks are normal in the 2SC and 3SC phases, 
there exist the corresponding normal currents 
$j^{r}_{\mu}(A)=-\Pi_{\mu\nu}^{el}A^{\nu}$ 
and $j^{r}_{\mu}(G_8 )=-\Pi_{\mu\nu}^{gl}G_{8}^{\nu}$ which 
however do not contribute in the static limit under consideration to the above 
Ginzburg - Landau equations, cf. \cite{BS93}. Thus, the qualitative behavior 
of the static magnetic field for all three 2SC, 3SC, and CFL phases is the 
same. 
 
We will solve the Ginzburg-Landau 
equations (\ref{ku}), (\ref{ka}), (\ref{kq}) for the case of a homogeneous 
external magnetic field for three types of superconducting regions: $a)$ a 
semi-infinite region with a planar boundary, $b)$ a cylindrical region and 
$c)$ a spherical region. 
The cases a) and c) simulate the behavior of the magnetic field 
in  quark cores of massive neutron stars. 
A discussion of all the three cases might be interesting in connection with 
the expectation that the slab, the rod and the droplet structures may exist 
within the mixed quark-hadron phase of the neutron stars, cf. \cite{16}, 
\cite{17}. 
 
We assume a sharp boundary between the quark and hadron matter since the 
diffusion boundary layer is thin, of the order of the size of the confinement 
radius ($\sim 0.2\div 0.4$ fm), and we suppose that  the coherence length 
$l_{\xi}=\sqrt{\gamma /(-2a)}$ is not less than this value and the magnetic 
and gluomagnetic field penetration depth $\lambda_q$ is somewhat larger 
than the confinement radius. 
Also we assume that the size of the quark region is much larger than all 
mentioned lengths. 
 
\section{ Solution of Ginzburg-Landau equations} 
 
\bigskip Let us rewrite equations (\ref{ka}) and (\ref{kq}) for a 
homogeneous superconducting matter region (being either a type I 
superconductor, or a type II superconductor for $H<H_{c1}$) in the following 
form 
\begin{equation} 
\lambda _q^{2\;}{\rm rot\;rot}\stackrel{\rightarrow }{A}+\sin ^2\alpha 
\stackrel{\rightarrow }{A}=\sin \alpha \cos \alpha 
\stackrel{\rightarrow }{G}_8,  \label{kc} 
\end{equation} 
\begin{equation} 
\lambda _q^2~{\rm rot}~{\rm rot}\stackrel{\rightarrow }{G}_8
+\cos ^2\alpha \stackrel{\rightarrow }{G}_8=\sin \alpha \cos \alpha 
\stackrel{\rightarrow }{A}.  
\label{kd} 
\end{equation} 
The field $\stackrel{\rightarrow }{G}_8$ is defined from (\ref{kc}) as follows 
\begin{equation} 
\stackrel{\rightarrow }{G}_8=
\frac{\lambda _q^2~{\rm rot}~{\rm rot}\stackrel{\rightarrow }{A}+\sin ^2\alpha 
\stackrel{\rightarrow }{A}}{\sin \alpha \cos \alpha }.  
\label{ke} 
\end{equation} 
From equations (\ref{kd}) and (\ref{ke}) we obtain the relation 
\begin{equation} 
{\rm rot}~{\rm rot}\stackrel{\rightarrow }{G}_8
=-\cot \alpha ~{\rm rot}~{\rm rot}\stackrel{\rightarrow }{A}.  
\label{kf} 
\end{equation} 
Substitution of 
$\stackrel{\rightarrow }{G}_8$ from (\ref{ke}) into (\ref{kf}) yields 
\begin{equation} 
{\rm rot}~{\rm rot}~(\lambda _q^2~{\rm rot}~{\rm rot}
\stackrel{\rightarrow }{A}+\stackrel{\rightarrow }{A})=0.  
\label{kg} 
\end{equation} 
Introducing the new function $\stackrel{\rightarrow }{M}$, 
\begin{equation} 
\stackrel{\rightarrow }{M}={\rm rot rot}\stackrel{\rightarrow }{A\;}~,  
\label{kh} 
\end{equation} 
we obtain 
\begin{equation} 
\lambda _q^{2~}{\rm rot rot}\stackrel{\rightarrow }{M}
+ \stackrel{\rightarrow }{M}\;\,=0.  
\label{kl} 
\end{equation} 
Thus the vector potential $\stackrel{\rightarrow }{A}$ can be determined by 
simultaneous solution of equations (\ref{kh}) and (\ref{kl}), 
whereas the gluonic potential $\stackrel{\rightarrow }{G}_8$ is found from 
(\ref{ke}). 
 
For the solution of equations (\ref{ke}), (\ref{kh}) and (\ref{kl}) we also 
need appropriate boundary conditions. At the quark-hadronic matter boundary we 
require the continuity of the magnetic field and the vanishing of the gluon 
potential ($\stackrel{\rightarrow }{G}_8=0)$ due to gluon confinement.  
Also the potential $G_8$ and the magnetic induction can't be infinite 
within the region of their existence. 
As we shall see below, these conditions are sufficient for a unique 
determination of the magnetic and gluomagnetic fields inside the quark matter. 
 
Equations (\ref{ku}), (\ref{ka}) and (\ref{kq}) have an obvious solution for a 
homogeneous $(\nabla d_p=0)$ and infinite superconductor 
($\stackrel{\rightarrow }{A\;}=0,\;\,\stackrel{\rightarrow }{G}_8=0$) 
\begin{equation} 
d^2=-a/\beta >0. \label{km} 
\end{equation} 
This solution motivates the possibility of  existence of the 
complete Meissner effect for both the fields 
$\stackrel{\rightarrow }{A\;}$ and $\stackrel{\rightarrow }{G}_8$ 
inside the quark superconductor. 
It corresponds to the absolute minimum value of the free energy 
$f=f_n-a^2/(2\beta )$ , where $f_n$ is the free energy of the normal quark 
matter \cite{10}, \cite{13}. 
The presence of the fields 
$\;\stackrel{\rightarrow }{A\;}\neq 0,\;\,\stackrel{\rightarrow }{G}_8 \neq 0$ 
inside the quark region would increase the free energy. 
 
As we have mentioned, the estimates \cite{13}, \cite{15} have demonstrated 
that color superconductors are type II superconductors. 
Indeed, the Ginzburg - Landau parameter $\kappa =\lambda_q /l_{\xi} = 
\sqrt{\beta}/(\gamma q)> 3$. 
For type II superconductors one can drop the fields 
$\stackrel{\rightarrow }{A\;}$ and $\stackrel{\rightarrow }{G}_8$ 
in the Ginzburg - Landau equation (\ref{ku}) arriving at the solution 
(\ref{km}), since the coherence length $l_{\xi}$ is smaller than the 
penetration depth of the magnetic and gluomagnetic fields $\lambda_q$. 
Then in equations of motion (\ref{ka}) and (\ref{kq}) 
the penetration depth of the magnetic and gluomagnetic fields $\lambda_q$ can 
be put constant. 
This simplifies solution of the Ginzburg - Landau equations very much. 
For further analytical treatment of the problem we will assume that we deal 
with a type II superconductor although our main conclusion on the existence 
of the Meissner effect is quite general. 
 
\subsection{Planar boundary} 
 
Let us first consider a semi-infinite region of superconducting quark 
matter for $x<0$ with a planar boundary, which coincides with the $zy$ plane. 
The external static homogeneous magnetic field $H$ is directed along the 
$z$-axis, the external vector potential $\stackrel{\rightarrow }{A}$ is 
aligned in the $y$ direction $A_y (x)=Hx$. 
Then the internal potentials $\stackrel{\rightarrow }{A}$ and 
$\stackrel{\rightarrow }{G}_8$ also have only $y$ components: $A_y=A_y(x)$ 
and $G_{8y}=G_{8y}(x)$, $\mbox{div} \stackrel{\rightarrow }{A\;}=0$, 
$\mbox{div} \stackrel{\rightarrow}{G}_8 =0$. 
Simultaneous solution of equations (\ref{kh}) and (\ref{kl}) determines the 
electromagnetic vector potential \thinspace $A_y\;$as follows 
\begin{equation} 
A_y=c_1\exp (\frac x{\lambda _q})+c_2x+c_3.  
\label{kn} 
\end{equation} 
We put $c_2=0$ because otherwise 
$A_y\rightarrow \infty $ and $G_{8y}\rightarrow \infty $ for 
$x\rightarrow \infty $ that would lead 
to a complete destruction of the condensate. 
We search for an energetically favorable unique solution of the problem 
satisfying the above mentioned boundary conditions. 
Thus we further put 
\begin{equation} 
A_y=c_1\exp (\frac x{\lambda _q})+c_3.  
\label{kp} 
\end{equation} 
Substitution of  the solution (\ref{kp}) into equation (\ref{ke}) yields 
the gluonic potential $G_{8y}$, 
\begin{equation} 
G_{8y}=-\cot \alpha \;c\,_1\exp (\frac x{\lambda_q})+\tan \alpha \;c_3\,. 
\label{kt} 
\end{equation} 
The boundary condition $G_{8y}(x=0)=0$ determines the constant 
$c_3=c_1\cot ^2\alpha $ . 
For the vector potential of the magnetic field we obtain 
\begin{equation} 
A_y=c_1(\exp (\frac x{\lambda _q})+\cot ^2\alpha ).  
\label{kr} 
\end{equation} 
We can determine $c_1$ from the remaining boundary condition $dA_y(x=0)/dx=0$ 
which yields $c_1=H\lambda _q$ . 
Finally the potentials render 
\begin{equation} 
A_y=H\lambda _q(\exp (\frac x{\lambda _q})+\cot ^2\alpha )~, 
\label{ks} 
\end{equation} 
\begin{equation} 
G_{8y}=-\cot \alpha \;H\,\lambda _q(\exp (\frac x{\lambda _q})-1)~. 
\label{kx} 
\end{equation} 
Then, for the magnetic induction\ $B_z=dA_y/dx$ and for the gluonic field 
$\;K_z=dG_{8y}/dx~$ inside the quark superconductor we have the 
following expressions
\begin{equation} 
\stackrel{\rightarrow }{B}=
H\exp (\frac x{\lambda _q})~\stackrel{\rightarrow }{e}_z~,  
\label{ky} 
\end{equation} 
\begin{equation} 
\stackrel{\rightarrow }{K}=-\cot \alpha \;B\,~\stackrel{\rightarrow }{e}_z~.  
\label{kz} 
\end{equation} 
Ref. \cite{14} introduced 
the ''rotated'' fields $\stackrel{\rightarrow }{B^x}$ and 
$\stackrel{\rightarrow }{B^y}$, 
\begin{equation} 
\stackrel{\rightarrow }{B^x}=
-\sin \alpha \;\stackrel{\rightarrow }{B}~
+\cos \alpha~\stackrel{\rightarrow }{K},~  
\label{a} 
\end{equation} 
\begin{equation} 
\stackrel{\rightarrow }{B^y}=\cos \alpha ~\stackrel{\rightarrow }{B} 
+\sin \alpha \;\stackrel{\rightarrow }{K\,}.  
\label{b} 
\end{equation} 
It is easy to see that  solutions (\ref{ky}) and (\ref{kz}) yield $B_z^y=0$ 
and $B_z^x=-(H/{\sin \alpha })\exp (x/{\lambda _q}).$ 
Therefore there is no $\stackrel{\rightarrow }{B^y}$ field inside the 
superconducting quark matter and the $B_z^x$ field is expelled from the color 
superconductor. 
Consequently, in contradiction with the statement of Ref. \cite{14} 
there is a Meissner effect for the color superconductors. 
 
\subsection{Cylindrical structures} 
 
Now we shall consider  a cylindrical region of superconducting quark matter 
of radius $a$ , whose axis coincides with $z$ axis of cylindrical 
coordinates. 
Such a situation may occur in the mixed phase where the rods of quark matter imbedded 
in the hadron matter are 
possible configurations. 
The external homogeneous magnetic field $H$ is directed along 
the $z$ axis, the external vector potential $A$ is aligned in the $\varphi $ - 
direction $A_\varphi =Hr/2$ . The internal vector potentials 
$\stackrel{\rightarrow }{A}$ and $\stackrel{\rightarrow }{G}_8$ 
have only $\varphi $ - components $A_\varphi (r)$ and $G_{8\varphi }(r)$ . 
Thus, equation (\ref{kl}) acquires the form 
\begin{equation} 
\frac{d^2M_\varphi }{dr^2}+
\frac 1r\frac{dM_\varphi }{dr}-(\frac 1{r^2}+\frac 1{\lambda ^2})M_\varphi=0~. 
\label{c} 
\end{equation} 
The solution of this equation is $M_\varphi =-c_1I_1(r/\lambda _q),$ where 
$I_1(r/\lambda _q)$ is  the corresponding modified Bessel function. 
Consequently, 
equation (\ref{kh}) can be written in the following form 
\begin{equation} 
\frac{d^2A_\varphi }{dr^2}+\frac 1r\frac{dA_\varphi }{dr}-
\frac{A_\varphi }{r^2}=c_1I_1(\frac r{\lambda _q})~.  
\label{d} 
\end{equation} 
For the vector potential $A_\varphi $ we obtain 
\begin{equation} 
A_\varphi (r)=c_1\lambda _q^2I_1(\frac r{\lambda _q})+c_2r.  
\label{e} 
\end{equation} 
Then for the potential $G_{8\varphi }$ from (\ref{ke}) we find the following 
expression 
\begin{equation} 
G_{8\varphi }(r)=-\cot \alpha \;c_{1\,}\lambda _q^2I_1(\frac r{\lambda 
_q})+\tan \alpha \;c_2r.  
\label{f} 
\end{equation} 
The magnetic induction $B_z$ inside the superconductor is given by 
\begin{equation} 
B_z(r)=c_1\lambda _qI_0(\frac r{\lambda _q})+2c_2.  \label{g} 
\end{equation} 
We determine $c_2$ from the boundary condition $G_{8\varphi }(a)=0$ and $c_1$ 
from $B_z(a)=H$. Consequently,  final expressions for the potentials are 
\begin{equation} 
A_\varphi (r)=\frac{H\lambda _q}{P(a/\lambda _q,\alpha )}\left[ I_1(\frac 
r{\lambda _q})+\cot ^2\alpha \;\frac raI_1(\frac a{\lambda _q})\right] , 
\label{h} 
\end{equation} 
\begin{equation} 
G_{8\varphi }(r)=-\frac{H\lambda _q\cot \alpha }{P(a/\lambda _q,\alpha )} 
\left[ I_1(\frac r{\lambda _q})-\frac raI_1(\frac a{\lambda _q})\right] , 
\label{m} 
\end{equation} 
and  for the corresponding  fields we get
\begin{equation} 
B_z(r)=\frac H{P(a/\lambda _q,\alpha )}\left[ I_0(\frac r{\lambda _q})
+2\cot^2\alpha \;\frac{\lambda _q}aI_1(\frac a{\lambda _q})\right] ,  
\label{n} 
\end{equation} 
\begin{equation} 
K_z(r)=-\frac{H\cot \alpha }{P(a/\lambda _q,\alpha )}\left[ I_0(\frac 
r{\lambda _q})-2\frac{\lambda _q}aI_1(\frac a{\lambda _q})\right] , 
\label{l} 
\end{equation} 
where $P(a/\lambda _q,\alpha )$ is given by 
\begin{equation} 
P(a/\lambda _q,\alpha )=I_0(\frac a{\lambda _q})+2\frac{\lambda _q}a\cot 
^2\alpha \;I_1(\frac a{\lambda _q})~.  
\label{p} 
\end{equation} 
Thus, we obtain the following final expressions for the rotated fields 
\begin{equation} 
B_z^y=\frac{2\lambda _qH\cot \alpha }{aP(a/\lambda _q,\alpha )\sin \alpha }
I_1\left( \frac a{\lambda _q}\right) ,  
\label{q} 
\end{equation} 
\begin{equation} 
B_z^x=-\frac H{P(a/\lambda _q)\sin \alpha }I_0(\frac r{\lambda _q}). 
\label{r} 
\end{equation} 
We notice that the field $\stackrel{\rightarrow }{B^y}$ is homogeneous 
inside the quark matter. 
Thus one may expect that in presence of the magnetic field the slab structures 
are energetically favorable starting with a smaller quark fraction volume than 
in absence of the magnetic field since, as we have argued, the magnetic field 
is expelled from the slabs and it penetrates the cylindres. 
In application to the description of the quark core of the neutron star, the
discussion of the cylindrical case has no meaning and 
one should further consider the case of a spherical geometry. 
 
\bigskip 
 
\subsection{Spherical case} 
 
We now assume that the neutron star possesses a spherical core of radius $a$ 
consisting of color superconducting quark matter. 
The applied homogeneous magnetic field $H$ is directed along the z axis. 
The functions $\stackrel{\rightarrow }{M}$, $\stackrel{\rightarrow }{A}$ and 
$\stackrel{\rightarrow }{G_8}$ have only $\varphi $ - components: 
$M_\varphi (r,\vartheta ),$ $A_\varphi (r,\vartheta )$ and 
$G_{8\varphi }(r,\vartheta )$. 
For the solution of the equation (\ref{kl}) we use the ansatz $M_\varphi 
(r,\vartheta )=f(r)\sin \vartheta $. Then the equation (\ref{kl}) can be 
written in spherical coordinates as 
\begin{equation} 
\frac{d^2f}{dr^2}+\frac 2r\frac{df}{dr}-\left( \frac 2{r^2}+\frac 1{\lambda 
_q^2}\right) f=0.  
\label{s} 
\end{equation} 
The solution of equation (\ref{s}) which tends to zero at the centre of the 
core renders 
\begin{equation} 
f(r)=-\frac D{r^2}J(\frac r{\lambda _q}),  
\label{t} 
\end{equation} 
where 
\begin{equation} 
J(\frac r{\lambda _q})=\sinh (\frac r{\lambda _q})-\frac r{\lambda _q}\cosh 
(\frac r{\lambda _q}).  
\label{x} 
\end{equation} 
For the solution of equation (\ref{kh}) we use the ansatz $A_\varphi 
(r,\vartheta )=g(r)\sin \vartheta .$ Then equation (\ref{kh} ) acquires the 
form 
\begin{equation} 
\frac{d^2g}{dr^2}+\frac 2r\frac{dg}{dr}-\frac{2g}{r^2}
=\frac D{r^2}J(\frac r{\lambda _q}).  
\label{y} 
\end{equation} 
The general solution of equation (\ref{kh}) in spherical coordinates is 
given by  
\begin{equation} 
A_\varphi (r,\vartheta )
= \frac{D\lambda _q^2}{r^2}J(\frac r{\lambda _q})\sin \vartheta 
+ c_1r\sin \vartheta .  
\label{z} 
\end{equation} 
We also find the general solution for the gluonic potential from (\ref{ke}) 
\begin{equation} 
G_{8\varphi }(r,\vartheta )=-\cot \alpha \frac{D\lambda _q^2}{r^2}J(\frac 
r{\lambda _q})\sin \vartheta +\tan \alpha \;c_1r\sin \vartheta .  
\label{aa} 
\end{equation} 
The constant $c_1$ is determined from the boundary condition 
$G_{8\varphi }(a,\vartheta )=0$ 
as $c_1=D\lambda _q^2\cot ^2\alpha J(a/\lambda _q)/a^3$. 
Then expressions for the vector potential $A_\varphi$ and for the gluonic 
potential $G_{8\varphi }$ are written as 
\begin{eqnarray} 
A_\varphi (r,\vartheta ) &=&\frac{D\lambda _q^2}{r^2}\left[ J(\frac 
r{\lambda _q})+\cot ^2\alpha \frac{r^3}{a^3}J(\frac a{\lambda _q})\right] 
\sin \vartheta ,  
\label{aaa} \\ 
G_{8\varphi }(r,\vartheta ) &=&
-\cot \alpha \frac{D\lambda _q^2}{r^2}\left[J(\frac r{\lambda _q})
-\frac{r^3}{a^3}J(\frac a{\lambda _q})\right] \sin \vartheta .  \nonumber 
\end{eqnarray} 
Therefore, the expressions for the radial component $B_{ir}$ and the 
tangential component $B_{i\vartheta }$ acquire the form
\begin{eqnarray} 
B_{ir}(r,\vartheta ) &=&\frac{2D\lambda _q^2}{a^3}\left[ \frac{a^3}{r^3} 
J(\frac r{\lambda _q})+\cot ^2\alpha \;J(\frac a{\lambda _q})\right] 
\cos \vartheta ,  
\label{ab} \\ 
B_{i\vartheta }(r,\vartheta ) &=&
\frac{D\lambda _q^2}{a^3}\left[ \frac{a^3}{r^3}J_1(\frac r{\lambda _q})
-2\cot ^2\alpha \;J(\frac a{\lambda _q})\right] \sin \vartheta~,  \nonumber 
\end{eqnarray} 
where 
\begin{equation} 
J_1(\frac r{\lambda _q})=
(1+\frac{r^2}{\lambda _q^2})\sinh (\frac r{\lambda_q})
-\frac r{\lambda _q}\cosh (\frac r{\lambda _q}).  
\label{ac} 
\end{equation} 
Accordingly, the expressions for the radial component $K_{ir}$ and the 
tangential component $K_{i\vartheta }$ of the internal gluomagnetic field 
render 
\begin{eqnarray} 
K_{ir}(r,\vartheta ) &=&
-\cot \alpha \frac{2D\lambda _q^2}{a^3}
\left[ \frac{a^3}{r^3}J(\frac r{\lambda _q})
-J(\frac a{\lambda _q})\right] \cos \vartheta,  
\label{ad} \\ 
K_{i\vartheta }(r,\vartheta ) &=&
-\cot \alpha \frac{D\lambda _q^2}{a^3}
\left[ \frac{a^3}{r^3}J_1(\frac r{\lambda _q})
+2J(\frac a{\lambda_q})\right] \sin \vartheta .  
\nonumber 
\end{eqnarray} 
The solutions of the Maxwell equations outside the sphere are
\begin{eqnarray} 
B_{er}(r,\vartheta ) &=&(H+\frac{2m}{r^3})\cos \vartheta ,  
\label{ae} \\ 
B_{e\vartheta }(r,\vartheta ) &=&(-H+\frac m{r^3})\sin \vartheta .  \nonumber 
\end{eqnarray} 
We can find $D$ and $m$ from the boundary conditions 
$B_{ir}(a,\vartheta )=B_{er}(a,\vartheta ),$ $B_{i\vartheta }(a,\vartheta 
)=B_{e\vartheta }(a,\vartheta ).$ 
Thus, we obtain 
\begin{eqnarray} 
D &=&-\frac{3Ha}{2N\sinh (a/\lambda _q)},  \label{af} \\ 
m &=&-\frac{Ha^3}{2N}\left[ 1+3\frac{\lambda _q^2}{a^2}-3\frac{\lambda _q} 
a\coth \frac a{\lambda _q}\right] ,  
\nonumber 
\end{eqnarray} 
where $N$ is given by 
\begin{equation} 
N=1-3\cot ^2\alpha \left[ \frac{\lambda _q^2}{a^2}-\frac{\lambda _q}a\coth 
\frac a{\lambda _q}\right] .  
\label{ag} 
\end{equation} 
The final expressions for the internal magnetic fields have the form 
\begin{eqnarray} 
B_{ir} &=&
-\frac{3H\lambda _q^2}{a^2N\sinh (a/\lambda _q)}
\left[ \frac{a^3}{r^3}J(\frac r{\lambda _q})
+\cot ^2\alpha J(\frac a{\lambda _q})\right] \cos \vartheta ,  
\label{ah} \\ 
B_{i\vartheta } &=&
-\frac{3H\lambda _q^2}{2a^2N\sinh (a/\lambda _q)}
\left[ \frac{a^3}{r^3}J_1(\frac r{\lambda _q})
-2\cot ^2\alpha J(\frac a{\lambda_q})\right] \sin \vartheta .  \nonumber 
\end{eqnarray} 
The corresponding internal components of gluomagnetic field are given by 
\begin{eqnarray} 
K_{ir} &=
&\frac{3H\lambda _q^2\cot \alpha }{a^2N\sinh (a/\lambda _q)}
\left[ \frac{a^3}{r^3}J(\frac r{\lambda _q})
-J(\frac a{\lambda _q})\right] \cos \vartheta ,  
\label{al} \\ 
K_{i\vartheta } &=&
\frac{3H\lambda _q^2\cot \alpha }{2a^2N\sinh (a/\lambda_{q)}}
\left[ \frac{a^3}{r^3}J_1(\frac r{\lambda _q})
+2J(\frac a{\lambda_q})\right] \sin \vartheta .  \nonumber 
\end{eqnarray} 
We obtain therefore the following internal components of the rotated 
sourceless field $\stackrel{\rightarrow }{B^y}$ 
\begin{eqnarray} 
B_{ir}^y &=&
\frac{3H\lambda _q^2\cot \alpha }{Na^2\sin \alpha }
\left[ \frac a{\lambda _q}\coth \frac a{\lambda _q}-1\right] \cos \vartheta ,  
\label{am} \\ 
B_{i\vartheta }^y &=&
-\frac{3H\lambda _q^2\cot \alpha }{Na^2\sin \alpha }
\left[ \frac a{\lambda _q}\coth \frac a{\lambda _q}-1\right] \sin \vartheta . 
\nonumber 
\end{eqnarray} 
It is to be noticed that the components of the rotated field 
$\stackrel{\rightarrow }{B^y}$ depend only on the polar coordinate $\vartheta.$
The internal components of the rotated massive field 
$\stackrel{\rightarrow }{B^x}$ are 
\begin{eqnarray} 
B_{ir}^x &=&\frac{3H\lambda _q^2a}{Nr^3\sin \alpha \sinh (a/\lambda _q)}
J(\frac r{\lambda _q})\cos \vartheta ,  
\label{an} \\ 
B_{i\vartheta }^x &=&\frac{3H\lambda _q^2a}{2Nr^3\sin \alpha \sinh 
(a/\lambda q)}J_1(\frac r{\lambda _q})\sin \vartheta .  \nonumber 
\end{eqnarray} 
In a neutron star, the radius of the superconducting quark core $a$ is much 
larger than the penetration depth $\lambda _q.$ In this limit the unbroken 
rotated field components take the form
\begin{eqnarray} 
B_{ir}^y &=&\frac{3H\cot \alpha }{\sin \alpha }\frac{\lambda _q}a\cos 
\vartheta ,  \label{ap} \\ 
B_{i\vartheta }^y &=&
-\frac{3H\cot \alpha }{\sin \alpha }\frac{\lambda _q}a\sin \vartheta.\nonumber 
\end{eqnarray} 
In the very same limit, for the components of the rotated massive field 
$\stackrel{\rightarrow }{B^x}$ at the surface of the quark core we obtain 
\begin{eqnarray} 
B_{ir}^x(a,\vartheta ) &=&
-\frac{3H}{\sin \alpha }\frac{\lambda _q}a\cos \vartheta ,  
\label{aq} \\ 
B_{i\vartheta }^x(a,\vartheta ) &=&\frac{3H}{\sin \alpha }\sin \vartheta . 
\nonumber 
\end{eqnarray} 
Therefore, the rotated massive field component at the surface of the quark 
core $B_{i\vartheta ~}^x(a,\vartheta )$ is larger than B$_{i\vartheta 
}^y$ by the factor $a/\lambda _q\cot \alpha $ and the $B_{ir}^x(a,\vartheta )$ 
and $B_{ir}^y$ components are of the same order of magnitude. 
We notice that there is Meissner effect for the rotated field $B^x.$ 
Let us now estimate the rotated field $\stackrel{\rightarrow }{B^y}$ inside 
the star. 
For values of the external field strength $H=10^{12}$ G, the penetration depth 
$\lambda _q=1.7$ fm, the mixing angle $\alpha =0.05$ and the radius of the 
quark core $a=1$ km we obtain $B^y=2\cdot$ $10^{-4}$ G. 
Thus, we can safely neglect the rotated unbroken field 
$\stackrel{\rightarrow }{B^y}$ . 
Therefore we again insist that there is a Meissner effect, and the applied 
external static magnetic field is thereby almost completely screened within 
the spherical geometry. 
Only a tiny fraction of the field can penetrate into the superconducting quark 
cores of the neutron stars. 
 
\section{Conclusion} 
 
We have investigated the behaviour of color superconducting quark matter in 
an external static homogeneous magnetic field and could show that color and 
electric Meissner currents exist. 
For this purpose we have solved the Ginzburg-Landau equations for three types 
of superconducting regions: a) a semi-infinite region with planar boundary, 
b) a cylindrical region and c) a spherical region. 
We have obtained analytic expressions for magnetic and gluomagnetic fields 
inside the quark matter for all three cases. 
In application to  the quark cores of  massive neutron stars we have shown 
that one can neglect the rotated field $\stackrel{\rightarrow }{B^y}$ inside 
the superconducting quark core since the Meissner currents effectively screen 
the applied external static magnetic field. 
These results confirm the physical situation discussed in \cite{15}. 
 
\subsection*{Acknowledgement} 
D.B. and D.S. acknowledge the support from DAAD for mutual visits and the 
hospitality extended to them at the partner Universities in Yerevan and 
Rostock, respectively. 
Discussions during the International Workshop on {\it Physics of Neutron Star 
Interiors} at the ECT* Trento, Italy, have been the starting point for this 
work. The authors are grateful to the ECT* for support and hospitality.


\bigskip 
 
\begin{thebibliography}{99} 
\bibitem{1}   M. Alford, K. Rajagopal, F. Wilczek, 
		Phys. Lett. {\bf B 422}, 247, 1998. 
 
\bibitem{2}   R. Rapp, T. Sch\"afer, E.V. Shuryak, M. Velkovsky, 
		Phys. Rev. Lett. {\bf 81}, 53, 1998. 
 
\bibitem{3}   G.W. Carter, D. Diakonov, Phys. Rev.{\bf \ D 60}, 016004, 1999. 
 
\bibitem{4}   D. Blaschke, C.D. Roberts, Nucl. Phys.{\bf \ A 642}, 197, 1998. 
 
\bibitem{5}   J.C.R. Bloch, C.D. Roberts, S.M. Schmidt, Phys. Rev.
		{\bf C 60}, 65208, 1999. 
 
\bibitem{6}   M. Alford, K. Rajagopal, F.Wilczek, 
		Nucl. Phys. {\bf B 357}, 443, 1999. 
 
\bibitem{7}   T. Sch\"afer, F. Wilczek, Phys. Rev. Lett. {\bf 82}, 3956, 1999. 
 
\bibitem{8}   M. Alford, J. Berges, K. Rajagopal, 
		Nucl. Phys. {\bf B 558}, 219, 1999. 
 
\bibitem{9}   D. Blaschke, T. Kl\"{a}hn, D.N.Voskresensky, 
		Astrophys. J. {\bf 533}, 406, 2000. 
\bibitem{9a}  D. Blaschke, H. Grigorian, D.N.Voskresensky, 
		Astron. \& Astrophys. (in press), [astro-ph/0009120], 2000. 
 
\bibitem{10}   D. Bailin, A. Love, Phys. Rep. {\bf 107}, 325, 1984. 
 
\bibitem{11}   K. Makashima, in: {\it The structure and evolution of neutron 
		stars}, edited by D. Pines, R. Tamagaki, S. Tsuruta, 
		(Addison-Wesley, New York, 1992), p. 86. 
 
\bibitem{12}   D. Bhattacharya, G. Srinivasan, in: {\it X-Ray Binaries}, 
		edited by W.H.G. Levin, J.van Paradijs, E.P.G. van den Heuvel, 
		(Cambridge University Press, 1995) p. 495. 
 
\bibitem{13}   D. Blaschke, D.M. Sedrakian, K.M. Shahabasyan, 
		Astron. \& Astrophys. {\bf 350}, L 47, 1999. 
 
\bibitem{14}   M. Alford, J. Berges, K. Rajagopal, 
		Nucl. Phys. {\bf B 571}, 269, 2000. 
 
\bibitem{15}   D. Blaschke, D.M. Sedrakian, nucl-th/0006038, 2000. 
 
\bibitem{BS93}   E. Braaten, D. Segel, Phys. Rev. {\bf D 48}, 1478,  1993. 
 
\bibitem{16}   N.K. Glendenning, Phys. Rev. {\bf C 52}, 2250, 1995. 
 
\bibitem{17}   H. Heiselberg, C.J. Pethick, E.F. Staubo, 
		Phys. Rev. Lett. {\bf 70}, 1355, 1993. 
\end{thebibliography}
\end{document}